\documentclass[a4paper,10pt]{article}
\usepackage{amssymb}
\usepackage{amstext}
\usepackage{amsmath}
\usepackage{amscd}
\usepackage{amsthm}
\usepackage{amsfonts}
\usepackage{enumerate}
\usepackage{latexsym}
\usepackage{mathrsfs}
\usepackage{mathtools}
\usepackage{color}

\makeatletter
 
 \@addtoreset{equation}{section}
\makeatother

\def\C{\mathbb C}
\def\N{\mathbb N}
\def\S{\mathbb S}
\def\R{\mathbb R}
\def\Z{\mathbb Z}

\def\B{\mathcal B}
\def\d{\mathrm d}

\def\e{\mathrm e}

\def\F{\mathscr F}
\def\H{\mathcal H}

\def\K{\mathcal K}
\def\SS{\mathcal S}

\def\U{\mathcal U}
\def\V{\mathcal V}

\def\i{i}

\def\arctanh{{\mathrm{arctanh}}}

\def\Hrond{\mathfrak{h}}
\def\ltwo{L^2}
\def\[{\left[}
\def\]{\right]}
\def\({\left(}
\def\){\right)}
\def\sgn{\mathop{\mathrm{sgn}}\nolimits}
\def\Hi{\mathsf H}
\def\Pv{{\mathrm{P.v.}}}
\def\l{\ell}
\def\HH{\mathscr H}
\def\bv{\boldsymbol{\varphi}}
\def\diag{\mathrm{diag}}
\def\bpm{\begin{pmatrix}}
\def\epm{\end{pmatrix}}
\def\bsm{\begin{smallmatrix}}
\def\esm{\end{smallmatrix}}

\newtheorem{Theorem}{Theorem}[section]
\newtheorem{Remark}[Theorem]{Remark}

\newtheorem{Lemma}[Theorem]{Proposition}

\begin{document}

\title{On some integral operators appearing in scattering theory,
and their resolutions}

\author{
S. Richard\footnote{Supported by the grant\emph{Topological invariants
through scattering theory and noncommutative geometry} from Nagoya University,
and by JSPS Grant-in-Aid for scientific research (C) no 18K03328, and on
leave of absence from Univ.~Lyon, Universit\'e Claude Bernard Lyon 1, CNRS UMR 5208,
Institut Camille Jordan, 43 blvd.~du 11 novembre 1918, F-69622 Villeurbanne cedex,
France.}, T. Umeda\footnote{Supported by  JSPS Grant-in-Aid for scientific research (C) no 18K03340.}}

\date{\small}
\maketitle
\vspace{-1cm}

\begin{quote}
\emph{
\begin{enumerate}
\item[] Graduate school of mathematics, Nagoya University,
Chikusa-ku,\\Nagoya 464-8602, Japan
\item[] Department of Mathematical Sciences, University of Hyogo, Shosha,
Himeji 671-2201, Japan
\item[]\emph{E-mail:} richard@math.nagoya-u.ac.jp, umeda@sci.u-hyogo.ac.jp
\end{enumerate}
}
\end{quote}

\begin{abstract}
We discuss a few integral operators
and provide expressions for them in terms
of smooth functions of some natural self-adjoint operators.
These operators appear
in the context of scattering theory,
but are independent of any perturbation theory.
The Hilbert transform, the Hankel transform,
and the finite interval Hilbert transform
are among the operators considered.
\end{abstract}

\smallskip



\section{Introduction}

Investigations on the wave operators in the context of scattering theory have a long history,
and several powerful technics have been developed for the proof of their existence and of their completeness.
More recently, properties of these operators in various spaces have been studied,
and the importance of these operators for non-linear problems has also been acknowledged.
A quick search on MathSciNet shows that the terms \emph{wave operator(s)} appear in the title of numerous
papers, confirming their importance in various fields of mathematics.

For the last decade, the wave operators have also played a key role for the search
of index theorems in scattering theory, as a tool linking the scattering part of a physical system
to its bound states. For such investigations, a very detailed understanding of the
wave operators has been necessary, and it is during such investigations that several
integral operators or singular integral operators have appeared, and that their resolutions in terms of smooth functions
of natural self-adjoint operators have been provided.
The present review paper is an attempt to gather some of the formulas obtained
during these investigations.

Singular integral operators are quite familiar to analysts, and refined estimates
have often been obtained directly from their kernels. However, for index theorems or for topological properties
these kernels are usually not so friendly: They can hardly fit into any algebraic framework.
For that reason, we have been looking for representations in which these singular integral operators
have a smoother appearance. It turns out that for several integral operators
this program has been successful, and suitable representations have been exhibited.
On the other hand, let us stress that even though these formulas are necessary for
a $C^*$-algebraic approach, it seems that for pure analysis they do not lead to any
new refined estimates.

Let us now be more precise for a few of these operators, and refer to the subsequent sections
for more information.
The Hilbert transform is certainly one of the most famous and ubiquitous singular integral operators.
Its explicit form in $L^2(\R)$ is recalled in \eqref{eq_def_Hi}.
A nice representation of this singular operator does not take place directly in $L^2(\R)$, but by
decomposing this Hilbert space into odd and even functions, then it becomes possible to obtain
an expression for the Hilbert transform in terms of the operators $\tanh(\pi A)$ and $\cosh(\pi A)$
where $A$ denotes the generator of dilations in $L^2(\R_+)$.
Such an expression is provided in Proposition \ref{lem_Hi}.
Similarly, the Hankel transform whose definition in $L^2(\R_+)$ is recalled in \eqref{def_Han}
is an integral operator whose kernel involves a Bessel function of the first kind.
This operator is not invariant under the dilation group, and so does the inversion operator
$J$ defined on $f\in L^2(\R_+)$ by $[Jf](x) = \frac{1}{x} f(\frac{1}{x})$.
However, the product of these two operators is invariant under the dilation group
and can be represented by a smooth function of the generator of this group, see
Proposition \ref{Prop_Hankel} for the details.

For several integral operators on $\R_+$ or on $\R^n$, the dilation group plays an important role,
as emphasized in Section \ref{sec_2}.
On the other hand, on a finite interval $(a,b)$ this group does not play any role (and is even not defined on such a space).
For singular operators on such an interval, the
notion of \emph{rescaled energy representation} can be introduced, and then tools
from natural operators on $\R$ can be exploited. This approach is developed in Section \ref{sec_3}.

As a conclusion, this short expository paper does not pretend to be exhaustive or self-contained.
Its (expected) interest lies on the collection of several integral operators
which can be represented by smooth functions of some natural self-adjoint operators.
It is not clear to the authors if a general theory will ever be built from
these examples, but gathering the known examples in a single place seemed
to be a useful preliminary step.

\section{Resolutions involving the dilation group}\label{sec_2}

The importance of the dilation group in scattering theory is well-known, at least since the seminal
work of Enss \cite{E81}. In this section we recall a few integral operators which appeared during our investigations
on the wave operators.
They share the common property of being diagonal in the spectral representation of the generator of the dilation group.
Before introducing these operators, we recall the action of this group in $L^2(\R^n)$ and in $L^2(\R_+)$.

In $L^2(\R^n)$ let us set $A$ for the self-adjoint generator of the unitary group of dilations,
namely for $f\in L^2(\R^n)$ and $x\in \R^n$~:
\begin{equation*}
[\e^{itA}f](x):=\e^{nt/2}f(\e^t x).
\end{equation*}
There also exists a representation in $L^2(\R_+)$ which is going to play an important role:
its action is given by $[\e^{itA}f](x):=\e^{t/2}f(\e^t x)$ for $f\in L^2(\R_+)$
and $x\in \R_+$. Note that we use the same notation for these generators independently of
the dimension, but this slight abuse of notation will not lead to any confusion.

\subsection{The Hilbert transform}

Let us start by recalling that the Hilbert transform is defined
for $f$ in the Schwartz space $\SS(\R)$
by the formula
\begin{equation}\label{eq_def_Hi}
[\Hi f](x) := \frac{1}{\pi}\Pv \int_\R\frac{1}{x-y}\;\!f(y)\;\! \d y
= -\frac{i}{\sqrt{2\pi}}\int_\R\e^{ikx} \;\!\hat{k}\;\! [\F f](k)\;\!\d k.
\end{equation}
Here $\Pv$ denotes the principal value, $\hat{k} := \frac{k}{|k|}$
for any $k\in \R^*$ and $\F f$ stands for the Fourier transform of $f$
defined by
\begin{equation}\label{eq_Fourier}
[\F f](k):= \frac{1}{\sqrt{2\pi}} \int_\R\e^{-ikx} f(x)\;\! \d x.
\end{equation}
It is well known that this formula extends to a bounded operator
in $L^2(\R)$, still denoted by $\Hi$.
In this Hilbert space, if we use the notation $X$ for the canonical self-adjoint operator
of multiplication by the variable, and $D$ for the self-adjoint realization of the operator
$-i\frac{\d}{\d x}$, then the Hilbert transform also satisfies the equality
$\Hi=-i \sgn(D)$, with $\sgn$ the usual sign function.

Based on the first expression provided in \eqref{eq_def_Hi} one easily observes that
this operator is invariant under the dilation group in $L^2(\R)$.
Since this group leaves the odd and even functions invariant,
one can further decompose the Hilbert space in order to get its irreducible
representations, and a more explicit formula for $\Hi$.
Thus, let us introduce
the even$\;\!$/$\;\!$odd representation of $L^2(\R)$.
Given any function $\rho$ on $\R$, we write $\rho_{\rm e}$ and $\rho_{\rm o}$ for the
even part and the odd part of $\rho$. We then introduce the unitary map
$$
\U:\ L^2(\R) \ni f \mapsto  \sqrt2
\begin{pmatrix}
f_{\rm e}\\
f_{\rm o}
\end{pmatrix}
\in L^2(\R_+;\C^2).
$$
Its adjoint is given on $\big(\begin{smallmatrix}
h_1\\
h_2
\end{smallmatrix}\big)\in L^2(\R_+;\C^2)$ and for $x\in \R$ by
$$
\big[\U^*
\big(\begin{smallmatrix}
h_1\\
h_2
\end{smallmatrix}\big)\big](x)
:=\frac1{\sqrt2}
\[h_1(|x|)+\sgn(x)h_2(|x|)\].
$$

A new representation of the Hilbert transform can now be stated.
We refer to \cite[Lem.~3]{KR08} for the initial proof, and to \cite[Lem.~2.1]{RT10}
for a presentation corresponding to the one introduced here.

\begin{Lemma}\label{lem_Hi}
The Hilbert transform $\Hi$ in $L^2(\R)$ satisfies the following equality:
\begin{equation*}
\U\;\! \Hi\;\!\U^*=-i \(\begin{matrix}
0 & \tanh(\pi A)-i\cosh(\pi A)^{-1}\\
\tanh(\pi A)+i\cosh(\pi A)^{-1} & 0
\end{matrix}\).
\end{equation*}
\end{Lemma}

Note that this representation emphasizes several properties of the Hilbert transform.
For example, it makes it clear that the Hilbert transform exchanges even and odd functions.
By taking the equality $\tanh^2 + \cosh^{-2}=1$, one also easily deduces that the norm
of $\Hi$ is equal to $1$. This latter property can certainly not be directly
deduced from the initial definition of the Hilbert transform based on the principle value.

The Hilbert transform appears quite naturally in scattering theory,
as emphasized for example in the seminal papers \cite{DF,W,Y}.
The explicit formula presented in Lemma \ref{lem_Hi} is used
for the wave operators in \cite[Thm.~1.2]{RT10}.
A slightly different version also appears in \cite[Eq.~(1)]{KR08}.

\subsection{The Hankel transform}

Let us recall that the Hankel transform is a transformation involving a Bessel function of the first kind.
More precisely, if $J_m$ denotes the Bessel function of the first kind with $m\in \C$ and $\Re(m)>-1$,
the Hankel transform $\H_m$ is defined on $f\in C_{\rm c}^\infty(\R_+)$ by
\begin{equation}\label{def_Han}
[\H_m f](x) = \int_{\R_+}\sqrt{xy}\;\! J_m(xy)\;\! f(y) \;\!\d y.
\end{equation}
Note that slightly different expressions also exist
for the Hankel transform, but this version is suitable for its representation in $L^2(\R_+)$.
Let us also introduce the unitary and self-adjoint operator
$J: L^2(\R_+)\to L^2(\R_+)$ defined for $f\in L^2(\R_+)$ and $x\in \R_+$ by
$$
[Jf](x) = \frac{1}{x} f\Big(\frac{1}{x}\Big).
$$

It is now easily observed that neither $\H_m$ nor $J$ are invariant under
the dilation group. However, the products $J\H_m$ and $\H_m J$ are invariant, and thus have
a representation in terms of the generator of dilations.
The following formulas have been obtained in \cite[Prop.~4.5]{DR17}
based on an earlier version available in \cite[Thm.~6.2]{BDG11}.
Note that in the statement the notation $\Gamma$ is used for the usual Gamma function.

\begin{Lemma}\label{Prop_Hankel}
For any $m\in \C$ with $\Re(m)>-1$
the map $\H_{m}$ continuously extends to a bounded invertible operator on $L^2(\R_+)$
satisfying $\H_{m} = \H_m^{-1}$.
In addition, the equalities
\begin{equation*}
J \H_m=\Xi_m(A) \quad \hbox{and} \quad \H_m J =\Xi_m(-A) ,\
\end{equation*}
hold with
\begin{equation*}
\Xi_m(t):= \e^{i\ln(2)t} \frac{\Gamma(\frac{m+1+i t}{2})}{\Gamma(\frac{m+1-i t}{2})}\ .
\end{equation*}
\end{Lemma}

Let us mention that the function $t\mapsto \Xi_m(t)$ has not a very interesting asymptotic behavior
for large $|t|$. However, by taking the asymptotic behavior of the Gamma function into account,
one can observe that the product of two such functions has a much better behavior, namely
for any $m,m'\in \C$ with $\Re(m)>-1$ and $\Re(m')>-1$ the map
$$
t\mapsto \Xi_m(-t)\Xi_{m'}(t)
$$
belongs to $C\big([-\infty,\infty]\big)$ and one has
$\Xi_m(\mp\infty)\Xi_{m'}(\pm\infty)
=\e^{\mp\i \frac{\pi}{2}(m-m')}$.

Note that such a product of two function $\Xi_m$ appears at least in two distinct contexts:
For the wave operators of Schr\"odinger operators with an inverse square potential,
see \cite[Thm.~6.2]{BDG11}, \cite[Eq.~(4.24)]{DR17} and also
\cite{IR1,IR2}, and for the wave operators of an Aharonov-Bohm system \cite[Prop.~11]{PR11}.
Let us also mention that additional formulas in terms of functions of $A$ have been found
in \cite[Thm.~12]{PR11} as a result of a transformation involving a Bessel function of the first kind
and a Hankel function of the first kind.

\begin{Remark}\label{rem_referee}
Let us still provide a general scheme for operators in $L^2(\R_+)$
which can be written in terms of the dilation group. If $K$ denotes
an integral operator with a kernel $K(\cdot,\cdot)$ satisfying for any $x,y,\lambda \in \R_+$
the relation
\begin{equation}\label{eq_inv}
K(\lambda x,\lambda y)= \frac{1}{\lambda}K(x,y),
\end{equation}
then this operator commutes with the dilation group.
As a consequence, $K$ can be rewritten as a function of the generator $A$
of the dilation group in $L^2(\R_+)$, and one has $K=\varphi(A)$ with $\varphi$ given by
$$
\varphi(t)= \int_{0}^\infty K(1,y) \;\!y^{-\frac{1}{2}+it} \d y.
$$
Note that this expression can be obtained by using the general formula
$$
\varphi(A)f = \frac{1}{\sqrt{2\pi}}\int_\R \check{\varphi}(t)\;\!\e^{-itA}f
$$
in conjunction with the homogeneity relation \eqref{eq_inv}.
\end{Remark}

\subsection{A 3 dimensional example}

When dealing with scattering theory for Schr\"odinger operators in $L^2(\R^3)$ one more operator
commuting with the generator of dilations appears quite naturally, see \cite[Sec.~3]{KR12}.
Let us set $\F$ for the Fourier transform in $\R^3$ defined for $f\in \SS(\R^3)$ and $k\in \R^3$ by
$$
[\F f](k)=\frac{1}{(2\pi)^{3/2}}\int_{\R^3}\e^{-ik\cdot x}\;\! f(x)\;\!\d x.
$$
Then we can define for $f \in \SS(\R^3)$ and $x \neq 0$ the integral operator
\begin{equation}\label{defdeT}
[Tf](x)=-i \frac{1}{\sqrt{2\pi}}\int_{\R_+}
\frac{e^{ i \kappa\;\!|x|}}{\kappa\;\!|x|}\;\![\F f](\kappa \;\!\hat{x})\;\!\kappa^2 \;\!\d \kappa
\end{equation}
where we have again used the notation $\hat{x}:=\frac{x}{|x|}\in \S^2$.
An easy computation shows that this operator is invariant under the action of the dilation group.
It is thus natural to express the operator $T$ in terms of the operator $A$.
In fact, the operator $T$ can be further reduced by decomposing the Hilbert space
with respect to the spherical harmonics.

Let us set $\Hrond:=L^2(\R_+, r^2\;\!\d r)$ and consider the spherical coordinates $(r,\omega) \in \R_+\times \S^2$.
For any $\ell \in \N = \{0,1,2,\ldots\}$ and $m\in \Z$ satisfying $-\ell\leq m \leq \ell$,
let $Y_{\ell m}$ denote the usual spherical harmonics. Then, by taking into account the completeness of the family
$\{Y_{\ell m}\}_{\ell \in \N, |m|\leq \ell}$ in $L^2(\S^2,\d \omega)$, one has the
canonical decomposition
\begin{equation}\label{decomposition}
L^2(\R^3) = \bigoplus_{\ell \in \N, |m|\leq \ell} \HH_{\ell m}\ ,
\end{equation}
where $\HH_{\ell m}=\{f \in L^2(\R^3) \mid f(r \omega)=g(r) Y_{\ell m}(\omega) \hbox{ a.e.~for some }g \in \Hrond\}$.
For fixed $\ell \in \N$ we denote by $\HH_\ell$ the subspace of $L^2(\R^3)$ given by $\bigoplus_{-\ell \leq m \leq \ell}\HH_{\ell m}$.
Let us finally observe that since the dilation group acts only on the radial coordinate,
its action is also reduced by the above decomposition.
In other terms, this group leaves each subspace $\HH_{\ell m}$ invariant.

As a final ingredient, let us recall that the Fourier transform $\F$ also
leaves the subspace $\HH_{\ell m}$ of $L^2(\R^3)$ invariant.
More precisely, for any $g \in C_{\rm c}^\infty(\R_+)$ and for $(\kappa,\omega) \in \R_+\times \S^2$ one has
\begin{equation}\label{surFl}
[\F (gY_{\ell m})](\kappa \omega) = (-i)^\l \;\!Y_{\l m}(\omega) \int_{\R_+}r^2\;\! \frac{J_{\l+1/2}(\kappa r)}{\sqrt{\kappa r}}g(r)\;\!\d r\ ,
\end{equation}
where $J_\nu$ denotes the Bessel function of the first kind. So, we naturally set $\F _\l: C_c^\infty(\R_+) \to \Hrond$ by the relation $\F (g Y_{\l m}) = \F_\l(g)Y_{\l m}$ (it is clear from \eqref{surFl} that this operator does not depend on $m$). Similarly to the Fourier transform in $L^2(\R^3)$, this operator  extends to a unitary operator from $\Hrond$ to $\Hrond$.

\begin{Remark}
If $\V: L^2(\R_+,r^2 \d r)\to L^2(\R_+,\d r)$ is the unitary map defined by $[\V f](r):=rf(r)$
for $f\in L^2(\R_+,r^2 \d r)$,
then the equality $\V \F_\l \V^* = (-i)^\ell \H_{\ell+1/2}$ holds, where the r.h.s.~corresponds
to the Hankel transform defined in \eqref{def_Han}.
\end{Remark}

By taking the previous two constructions into account, one readily
observes that the operator $T$ is reduced by the decomposition \eqref{decomposition}.
As a consequence one can look for a representation of the operator $T$
in terms of the dilation group in each subspace $\HH_{\ell m}$.
For that purpose, let us define for each $\ell \in \N$ the operator $T_\ell$ acting on any $g \in C_{\rm c}^\infty(\R_+)$ as
\begin{equation*}
[T_\ell \;\!g](r)=-i \frac{1}{\sqrt{2\pi}}\int_{\R_+}
\frac{e^{ i\kappa\;\!r}}{\kappa\;\!r}\;\! [\F_\ell \;\!g ](\kappa)\;\!\kappa^2\;\!\d \kappa\ .
\end{equation*}

The following statement has been proved in \cite[Prop.~3.1]{KR12}.

\begin{Lemma}\label{tralala}
The operator $T_\l$ extends continuously to the bounded operator $\varphi_\l(A)$ in $\HH_{\l m}$ with
$\varphi_\l\in C\big([-\infty,\infty]\big)$ given explicitly for every $x\in \R$ by
$$
\varphi_\l(x)=
\frac{1}{2}e^{-i\pi \l /2}
\frac{\Gamma\big(\frac{1}{2}(\l+3/2+ix)\big)}{\Gamma\big(\frac{1}{2}(\l+3/2-ix)\big)}  \frac{\Gamma\big(\frac{1}{2}(3/2-ix)\big)}{\Gamma\big(\frac{1}{2}(3/2+ix)\big)}  \big(1 + \tanh(\pi x)-i\cosh(\pi x)^{-1}\big)
$$
and satisfying $\varphi_\l(-\infty)=0$ and $\varphi_\l(\infty)=1$.
Furthermore, the operator $T$ defined in \eqref{defdeT} extends continuously to the operator
$\bv (A)\in \B\big(L^2(\R^3)\big)$ acting as $\varphi_\l(A)$ on $\HH_\l$.
\end{Lemma}

Let us finally observe that in the special case $\ell=0$, one simply gets
$$
\varphi_0(A)=\frac{1}{2}  \big(1 + \tanh(\pi A)-i\cosh(\pi A)^{-1}\big).
$$
This operator appears in particular in the expression for the wave operators in $\R^3$,
see \cite[Sec.~2.1]{KR06} and \cite[Thm.~1.1]{RT13}.
The same formula but for $1$-dimensional system  can be found for example in
\cite[Sec.~2.4]{KR06} and in \cite[Thm.~2.1]{R16}.
The adjoint of this operator also appears for $1$-dimensional system in \cite[Sec.~2.3]{KR06}
and in the expression for the wave operator in a periodic setting \cite[Thm.~5.7]{RT17}.
The related formula $\frac{1}{2}  \big(1 + \tanh(\pi A/2)\big)$ is used for systems in dimension $2$
as for example in \cite[Sec.~2.2]{KR06} or in \cite[Thm.~1.1]{RT13_2}.

\begin{Remark}
Let us mention that operators similar to the one presented in \eqref{defdeT}
appear quite often in the context of scattering theory, and then such expressions can
be reformulated in a way similar to the one presented in Proposition \ref{tralala}.
For example, such kernels can be exhibited from the asymptotic expansion
of the generalized eigenfunctions of the relativistic Schr\"odinger operators
in dimension $2$ \cite[Thm.~6.2]{UW} or in dimension $3$ \cite[Thm.~10.2]{U}.
\end{Remark}

\section{Resolutions in the rescaled energy representation}\label{sec_3}

In the previous section, the dilation group played a special role since all
operators were invariant under its action. For many other
singular kernels appearing in scattering theory, this is no more true,
and in many settings there is no analog of the dilation group.
However, a replacement for the operator $A$ can often be found
by looking at the \emph{rescaled energy representation}, see
for example \cite[Sec.~2.4]{BSB12} and \cite[Sec.~3.1]{SB16}.
The main idea in this approach is to rescale the underlying space
such that it covers $\R$, and then to use the canonical operators $X$ and $D$
on $\R$. Let us stress that here the \emph{energy space} corresponds
to the underlying space since the following operators are directly defined in the energy representation.

\subsection{The finite interval Hilbert transform}\label{sec_finite}

In this section we consider an analog of the Hilbert transform
but localized on a finite interval. More precisely, let us consider
the interval $\Lambda:=(a,b)\subset \R$.
For any $f\in C^\infty_{\rm c}(\Lambda)$ and $\lambda \in \Lambda$ we consider the operator
defined by
\begin{equation}\label{defT}
[Tf](\lambda):=\frac{1}{\pi} \;\Pv\!\!\int_\Lambda \frac{1}{\lambda - \mu} \;\!f(\mu)\;\!\d \mu.
\end{equation}
This operator corresponds to the Hilbert transform but restricted to a
finite interval.

In order to get a better understanding of this operator, let us
consider the Hilbert space $L^2(\R)$ and the unitary map $\U:L^2(\Lambda)\to L^2(\R)$
defined on any $f \in L^2(\Lambda)$ and for $x \in \R$ by
\begin{equation*}
[\U f](x):=\sqrt{\frac{b-a}{2}}\frac{1}{\cosh(x)}\;\!f\Big(
\frac{a+b\e^{2x}}{1+\e^{2x}}\Big)\ .
\end{equation*}
The adjoint of this map is given for $h \in L^2(\R)$ and $\lambda \in \Lambda$ by
\begin{equation*}
[\U^*h](\lambda) = \sqrt{\frac{b-a}{2}}\frac{1}{\sqrt{(\lambda -a)(b-\lambda)}}
\;\!h\Big(\frac{1}{2}\ln\frac{\lambda-a}{b-\lambda}\Big)\ .
\end{equation*}

Let us now denote by $L$ the operator of multiplication by the variable in $L^2(\Lambda)$
and set $\rho(L)$ for the operator of multiplication in $L^2(\Lambda)$ by a function
$\rho \in L^\infty(\Lambda)$.
Then, a straightforward computation leads to the following expression for its representation in $L^2(\R)$:
$\tilde \rho(X):= \U \;\!\rho(L)\;\!\U^*$ is the operator of multiplication by the function
$x\mapsto \tilde \rho(x)=\rho\big(\frac{a+b\e^{2x}}{1+\e^{2x}}\big)$.
In particular, by choosing $\rho(\lambda)= \lambda$ one obtains that $\U\;\! L\;\! \U^*$
is the operator $\tilde \rho(X)=\frac{a+b\e^{2X}}{1+\e^{2X}}$.
Note that the underlying function is strictly increasing on $\R$ and takes the asymptotic values
$\tilde \rho(-\infty)=a$ and $\tilde \rho(\infty) = b$.

Let us now perform a similar conjugation to the operator $T$.
A straightforward computation leads then to the following equality for any $h \in C^{\infty}_{\rm c}(\R)$ and $x \in \R$:
\begin{equation*}
[\U\;\! T\;\!\U^*h](x)=
\frac{1}{\pi} \;\Pv\!\!\int_\R \frac{1}{\sinh(x-y)}\;\!h(y)\;\d y \ .
\end{equation*}
Thus if we keep denoting by $D$ the self-adjoint operator corresponding to $-i\frac{\d}{\d x}$ in $L^2(\R)$,
and if one takes into account the formula
\begin{equation*}
\frac{i}{\pi}\;\Pv\!\! \int_\R\frac{\e^{-ixy}}{\sinh(y)}\;\d y= \tanh\big(\pi x/2\big)
\end{equation*}
one readily gets:

\begin{Lemma}
The following equality holds
\begin{equation}\label{defU}
\U\;\! T\;\! \U^* =-i\tanh\big(\pi D/2\big).
\end{equation}
\end{Lemma}

Such an operator plays a central role for the wave operator in the context
of the Friedrichs-Faddeev model \cite[Thm.~2]{IR12}.
Let us also emphasize one of the main interest of such a formula.
Recall that $X$ and $D$ satisfy the usual canonical commutation relations in $L^2(\R)$.
Obviously, the same property holds for the self-adjoint operators
$X_\Lambda:=\U^* \;\!X \;\!\U$ and $D_\Lambda:=\U^*\;\! D\;\! \U$ in $L^2(\Lambda)$.
More interestingly for us is that for any
functions $\varphi\in L^\infty(\R)$ and $\rho\in L^\infty(\Lambda)$
the operator $\varphi(D)\;\!\tilde \rho(X)$ in $L^2(\R)$ is unitarily equivalent to
the operator $\varphi(D_\Lambda)\;\!\rho(L)$ in $L^2(\Lambda)$.
In particular, this allows us to define quite naturally isomorphic $C^*$-subalgebras of
$\B\big(L^2(\R)\big)$ and of $\B\big(L^2(\Lambda)\big)$,
either generated by functions of $D$ and $X$, or by functions of $D_\Lambda$ and $L$.
By formula \eqref{defU}, one easily infers that the singular operator $T$ defined
in \eqref{defT} belongs to such an algebra.

\subsection{The finite interval Hilbert transform with weights}

The operator considered in this section is associated with a discrete adjacency operator on $\Z$.
Once considered in its energy representation, this operator leads to a Hilbert transform on a finite interval
multiplied by some weights.

We consider the Hilbert space $L^2\big((-2,2)\big)$ and the weight function
$\beta: (-2,2)\to \R$ given by
$$
\beta(\lambda):=\big(4-\lambda^2\big)^{1/4}.
$$
For any $f\in C_{\rm c}^\infty\big((-2,2)\big)$ and for $\lambda \in (-2,2)$ we define the operator
\begin{equation}\label{eq_Hi_weight}
[Tf](\lambda):=\frac{1}{2\pi i} \; \Pv \int_{-2}^2\beta(\lambda)\frac{1}{\lambda-\mu}\beta(\mu)^{-1} f(\mu)\;\!\d \mu.
\end{equation}
Clearly, this operator has several singularities: on the diagonal but also at $\pm 2$.

In order to get a better understanding of it, let us introduce the unitary
transformation $\U : L^2\big((-2,2)\big) \to L^2(\R)$ defined on $f \in L^2\big((-2,2)\big)$ by
$$
[\U f](x) := \sqrt{2}\frac{1}{\cosh(x)}f\big(2\tanh(x)\big).
$$
The action of its adjoint is given on $h\in \ltwo(\R)$ by
$$
[\U^*h](\lambda) = \frac{\sqrt{2}}{\sqrt{4-\lambda^2}}\;\!h\big(\arctanh(\lambda/2)\big).
$$
We also introduce the multiplication operators
$b_\pm(X)\in \B\big(\ltwo(\R)\big)$ defined by the real functions
$$
b_\pm(x):=\frac{\e^{x/2}\pm\e^{-x/2}}{(\e^x+\e^{-x})^{1/2}}.
$$
The function $b_+$ is continuous, bounded, non-vanishing, and satisfies
$\lim_{x\to \pm \infty}b_+(x)=1$.
The functions $b_-$ is also continuous, bounded, and satisfies
$\lim_{x\to \pm \infty}b_-(x)=\pm 1$.

With these notations, the following statement has been proved in \cite{NRT19}.

\begin{Lemma}
One has
\begin{equation}\label{eq_kernel}
\U \;\!T\;\!\U^*
 = -\frac{1}{2}\Big[b_+(X) \tanh(\pi D) b_+(X)^{-1}
- i b_-(X) \cosh(\pi D)^{-1} b_+(X)^{-1}\Big].
\end{equation}
\end{Lemma}

Note that a slightly simpler expression is also possible, once a compact error is accepted.
More precisely, since the functions appearing in the statement
of the previous proposition have limits at $\pm \infty$ the operator in the r.h.s.~of \eqref{eq_kernel} can be rewritten as
\begin{equation}\label{eq_mod_compact}
-\frac{1}{2}\Big[ \tanh(\pi D)
-i \tanh(X) \cosh(\pi D)^{-1}\Big] + K
\end{equation}
with $K\in \K\big(\ltwo(\R)\big)$, see for example \cite{Co} for a justification of the compactness of
the commutators.
Note that this expression can also be brought back to the initial representation
by a conjugation with the unitary operator $\U$.
Let us also mention that the operators obtained above play an important
role for the wave operator of discrete Schr\"odinger operators on $\Z^n$.
Such operators are currently under investigations in \cite{IT19} and in \cite{NRT19}.

\subsection{The upside down representation}

In this section we deal with a singular kernel which is related to a $1$-dimensional
Dirac operator. Compared to the operators introduced so far, its specificity
comes from its matrix-values. Dirac operators depend also on a parameter $m$ which we
choose strictly positive. The following construction takes already place in the
energy representation of the Dirac operator, namely on its spectrum.

Let us define the set
$$
\Sigma:= (-\infty,-m)\,\cup\,(m,+\infty)
$$
and for each $\lambda \in \Sigma$ the $2\times 2$ matrix
$$
B(\lambda) = \frac{1}{\sqrt{2}}\;\!
\diag\left(
\sqrt[4]{\frac{\lambda-m}{\lambda+m}}, \sqrt[4]{\frac{\lambda+m}{\lambda-m}}\right).
$$
Clearly, for any $\lambda\in \Sigma$ the matrix $B(\lambda)$ is well defined and invertible, but
it does not have a limit as $\lambda \searrow m$ or as $\lambda \nearrow -m$.
For $f\in C_{\rm c}^\infty(\Sigma;\C^2)$ we consider the singular operator
$T$ defined by
\begin{equation*}
[Tf](\lambda):=
\frac{1}{\pi}  B(\lambda)^{-1} \
\Pv \!\!\int_\Sigma \frac{1}{\lambda-\mu}\;\! B(\mu) \;\!f(\mu)\;\!\d \mu.
\end{equation*}

The trick for this singular operator is to consider the following unitary transformation
which sends the values $\pm m$ at $\pm \infty$,
while any neighbourhood of the points $\pm \infty$ is then located near the point $0$.
More precisely, let us define the unitary operator
$\U: L^2(\Sigma;\C^2) \to L^2(\R;\C^2)$ given for $f \in L^2(\Sigma;\C^2)$ and $x \in \R$ by
$$
\big[\U f\big](x) :=\sqrt{2m} \;\!\frac{\e^{x /2}}{\e^{x}-1} \, f\Big(m\frac{\e^{x}+1}{\e^{x}-1}\Big).
$$
The adjoint of the operator $\U$ is provided for $h \in L^2(\R;\C^2)$ and $\lambda \in \Sigma$ by the expression
$$
\big[\U^* h\big](\lambda) = \sqrt{2m} \;\!\sqrt{\frac{\lambda+m}{\lambda-m}}\;\!\frac{1}{\lambda+m}\;\!
\,h\Big(\ln\Big[\frac{\lambda+m}{\lambda-m}\Big]\Big).
$$
We shall now compute the kernel of the operator $\U \;\!T \;\!\U^*$, and observe that this new kernel has a very simple form.

For that purpose, we keep the notations $X$ and $D$ for the canonical self-adjoint operators on $L^2(\R)$,
and denote by $\F$ the Fourier transform in $L^2(\R;\C^2)$, namely two copies of the Fourier transform \eqref{eq_Fourier}.
One then checks, by a direct computation, that for any measurable function $\rho:\Sigma\to M_2(\C)$ one has
\begin{equation*}
\U \;\!\rho(L)\;\!\U^*=\rho\Bigg(m\frac{\e^X+1}{\e^X-1}\Bigg).
\end{equation*}
Furthermore, for any $f=(f_1,f_2) \in C_{\rm c}^\infty(\R;\C^2)$ and $x \in \R$,
it can be obtained straightforwardly that
\begin{align*}
& [\U \;\!T \;\!\U^* h](x) \\
& = \frac{1}{4\pi} \Pv \!\!\int_\R
\bpm
\frac{1}{\sinh((y-x)/4)} - \frac{1}{\cosh((y-x)/4)} & 0 \\
0 & \frac{1}{\sinh((y-x)/4)} + \frac{1}{\cosh((y-x)/4)}
\epm
h(y)\;\!\d y .
\end{align*}

By summing up the information obtained so far one obtains:

\begin{Lemma}
For any $m> 0$ one has

\begin{equation*}
\U \;\!T\;\! \U^*=i
\bpm
\tanh(2\pi D) + i\cosh(2\pi D)^{-1} & 0 \\
0 & \tanh(2\pi D) -i \cosh(2\pi D)^{-1}
\epm .
\end{equation*}
\end{Lemma}

We refer to \cite[Sec.~III.D]{PR14} for the details of the computation, and for the use of this
expression in the context of $1$-dimensional Dirac operators.
Note that in Section IV of this reference the $C^*$-algebraic
properties mentioned at the end of Section \ref{sec_finite} are exploited
and the construction leads naturally to some index theorem in scattering theory.

\subsection*{Acknowledgement}
S. Richard thanks the Department of Mathematics of the National University of Singapore
for its hospitality in February 2019.
The authors also thank the referee for suggesting the addition of Remark \ref{rem_referee}.
Its content is due to him/her.

\end{document}